\documentclass[a4paper]{article}

\usepackage[english]{babel}
\usepackage[utf8x]{inputenc}
\usepackage[T1]{fontenc}
\usepackage{comment}
\usepackage{fullpage}

\usepackage{amsmath}
\usepackage{amsfonts}
\usepackage{graphicx}
\usepackage{orcidlink}
\usepackage{authblk}
\usepackage{matlab-prettifier}
\usepackage{listings}
\usepackage{color}

\newtheorem{theorem}{Theorem}
\newtheorem{proposition}[theorem]{Proposition}
\newtheorem{remark}[theorem]{Remark}

\definecolor{dkgreen}{rgb}{0,0.6,0}
\definecolor{gray}{rgb}{0.5,0.5,0.5}
\definecolor{mauve}{rgb}{0.58,0,0.82}
\definecolor{codegreen}{rgb}{0,0.6,0}
\definecolor{codegray}{rgb}{0.5,0.5,0.5}
\definecolor{codepurple}{rgb}{0.58,0,0.82}
\definecolor{mygreen}{RGB}{28,172,0}
\definecolor{mylilas}{RGB}{170,55,241}
\definecolor{backcolour}{rgb}{0.95,0.95,0.92}

\lstdefinestyle{Rstyle}{
  frame=tb,
  language=R,
  aboveskip=3mm,
  belowskip=3mm,
  showstringspaces=false,
  columns=flexible,
  basicstyle={\small\ttfamily},
  numbers=none,
  numberstyle=\tiny\color{gray},
  keywordstyle=\color{blue},
  commentstyle=\color{dkgreen},
  stringstyle=\color{mauve},
  breaklines=true,
  breakatwhitespace=true,
  tabsize=3
}

\lstdefinestyle{Matlabstyle}{
  style=Matlab-editor,
  basicstyle=\mlttfamily,
  escapechar=`,
}

\title{DEViaN-LM: An \texttt{R} Package for Detecting Abnormal Values in the Gaussian Linear Model}

\author[1,2]{Geoffroy Berthelot~\orcidlink{0000-0003-4036-6114}}
\author[1]{Guillaume Sauli\`{e}re~\orcidlink{0000-0001-8263-6456}}
\author[3]{J\'{e}r\^{o}me Dedecker~\orcidlink{0000-0002-8838-0356}}
\affil[1]{Institut de Recherche bioM\'{e}dicale et d'Epid\'{e}miologie du Sport (IRMES, UPR7329), INSEP (Institut National du Sport, de l'Expertise et de la Performance), Paris, France}
\affil[2]{Research Laboratory for Interdisciplinary Studies (RELAIS), 89100 Sens, France}
\affil[3]{Universit\'{e} Paris Cit\'{e}, CNRS, UMR 8145, MAP5, F-75006 Paris, France}
\date{}

\begin{document}

\maketitle

\begin{abstract}
The \textbf{DEViaN-LM} is a \textbf{R} package that allows to detect the values poorly explained by a Gaussian linear model. The procedure is based on the maximum of the absolute value of the studentized residuals, which is a free statistic of the parameters of the model. This approach makes it possible to generalize several procedures used to detect abnormal values during longitudinal monitoring of certain biological markers. In this article, we describe the method used, and we show how to implement it on different real datasets.
\end{abstract}

\section{Introduction} \label{sec:intro}
The detection of outliers or abnormal values in a series of observations is one of the core problems in many applied fields. These abnormal values may represent errors or events of interest such as a novelty. Outliers detection finds application in different areas including, but not limited to: fraud detection, activity monitoring, fault diagnosis, structural defect detection, satellite image analysis, medical condition monitoring, pharmaceutical research, detection of unexpected entries in database or mislabeled data in a training data set \cite{hodge2004survey}.

In medicine, detecting abnormal values is typically useful in the biological monitoring of individuals, where blood or urine markers are collected on a temporal basis. Usually, biological values of a given individual are collected and compared to the statistical parameters of a population, which is deemed representative of the biological characteristics of the tested individual. However, \cite{malcovati2003hematologic} and \cite{egger2016interindividual} underlined that the inter-individual variability is higher than the intra-individual one, suggesting that the use and design of population data for estimating intra-individual variability is an issue. In line with \cite{sauliere2019z}, our approach rather aims at detecting if some observations are abnormal, based on the characteristics of the individual only. This is particularly relevant in the context of personalized/precision medicine which favors an individualized approach: ``Precision Medicine refers to the tailoring of medical treatment to the individual characteristics of each patient'' \cite{national2011toward}.

Let us describes the simplest procedure, based on a Z-score, to detect if the observation $x_n$ is  abnormal, when $x_n$ is the last observation  from a sequence of independent and identically distributed (iid) Gaussian random variables $X_1, \ldots , X_n$. In that case, the Z-score statistic is simply defined by
$$
    ZS_n = \dfrac{X_n - \bar{X}_{n-1}}{\hat{\sigma}_{n-1} \sqrt{1+\dfrac{1}{n-1}}}
$$
where
$$
    \bar{X}_{n-1} = \dfrac{1}{n-1} \sum_{k=1}^{n-1} X_k  \quad\quad\quad  \hat{\sigma}_{n-1}^2 = \dfrac{1}{n-2} \sum_{k=1}^{n-1} \left( X_k - \bar{X}_{n-1} \right)^2
$$
The Z-score provides an intuitive mechanism to determine the magnitude by which the observation $x_n$ diverges from the others in a sequence. If this divergence is large enough, the observation can be deemed as an outlier. This procedure works well, because in this Gaussian context, the distribution of $ZS_n$ is the Student($n-2$) distribution; in particular, it does not depends on the unknown quantities $\mu ={\mathbb E}(X_i)$ and $\sigma^2=\text{Var}(X_i)$.

The Z-score has found applications in biology (\cite{cheadle2003analysis}), neuroimaging (\cite{ishii2000diagnostic}), psychology (\cite{guilford1973structure}), sport (\cite{sauliere2019z}) and medicine (\cite{dallaire2015bias}) among others. Let us note that $ZS_n$ is in fact the $n$-th studentized residuals in the simple Gaussian linear model $X_i= \mu + \varepsilon_i$, where $\varepsilon_1, \ldots, \varepsilon_n$ is iid with ${\mathcal N}(0, \sigma^2)$ marginal distribution. Starting from this simple fact, \cite{sauliere2019z} have generalized this approach to the problem of detecting abnormal values in the series $x_1, \ldots, x_n$;  they proposed to use as a statistic the maximum of the absolute value of the studentized residuals, providing a table of the quantiles of this statistic.

In this paper, we generalize the approach of \cite{sauliere2019z} to the Gaussian linear model. As a consequence of Proposition \ref{main} below, we get that the distribution of the maximum of the absolute value of the studentized residuals does not depends of the unknown parameters of the linear model. However, it depends on the design matrix, so that a universal quantiles table cannot be used. Our second  objective in then to propose the \textbf{DEViaN-LM} package (Detecting  Extremal Values in a Normal Linear Model) that allows, for any given design matrix, to estimate the quantiles of this statistic via Monte-Carlo simulations, and thus to detect all the abnormal observations for a given risk level.

The paper is organized as follows: in Section 2 we describes in details the statistical procedure, an we prove the main result (Proposition \ref{main}). In Section 3 we present some applications of the \textbf{DEViaN-LM} package to some real data sets. In Section 4 we show that \textbf{DEViaN-LM} has been time-optimized, comparing it to direct procedures using \textbf{R} or \textbf{MATLAB} functions.

\section{Detecting abnormal values from studentized residuals}
We consider the homoscedastic linear model
\begin{equation}\label{mod}
X=M \theta + \varepsilon
\end{equation}
where $M$ is an $n \times p$  random matrix, $\varepsilon=( \varepsilon_1, \ldots, \varepsilon_n)'$ is a Gaussian random vector independent of $M$, whose coordinates are independent and identically distributed (with mean 0 and variance $\sigma^2$), and $\theta=(\theta_1, \ldots , \theta_p)'$ is the unknown vector of parameters.

One observes the random vector $X=(X_1, \, \ldots , X_n)'$ and the random design $M$. In this context we want to detect ``abnormal observations'', that is observations $x_i$'s that are poorly explained by the model.

Before describing our method in more details, let us mention that we shall work conditionally on the design matrix $M$, which is a standard approach when dealing with Model \eqref{mod}. Consequently, for the sake of simplicity and without loss of generality, we assume from now that $M$ is fixed.

Assume that  $n>p+1$ and that the columns of $M$ are linearly independent (so that $M'M$ is invertible). Let $M_{(i)}$ be the matrix $M$ deprived of its $i$-th row, and $X_{(i)}$ be the vector $X$ deprived of its $i$-th coordinate, and assume that for any $i \in \{1, \ldots, n\}$ the matrix $M_{(i)}' M_{(i)}$ is invertible. Let now $\hat \theta_{n, i}$ be the least squares estimator of $\theta$ based on the variables $X_j$'s, for $j\neq i$:
$$
    \hat \theta_{n, i}= \left(M_{(i)}' M_{(i)}\right)^{-1} M'_{(i)} X_{(i)}
$$
Let $L_i$ be the $i$-th row of the matrix $M$, and let $\hat X_{n,i}$ be the prediction of $X_i$ obtained from $\hat \theta_{n, i}$, that is $\hat X_{n,i}=L_i \hat \theta_{n, i}$. Standard computations show that
$$
    \mathrm{Var}(X_i-\hat X_{n,i})= \sigma^2 \left ( 1 + L_i (M'_{(i)} M_{(i)})^{-1}L_i'\right).
$$
Let
$$
    \hat \sigma_{n, i}^2= \frac 1 {n-p-1}
    \sum_{k \in \{1, \dots, n\}, k \neq i}
    \left(X_k-L_k \hat \theta_{n, i}\right)^2
$$
be the estimator of $\sigma^2$ based on the variables $X_j$'s, for $j\neq i$. Define then
$$
    \hat e_i(X)=\frac {X_{i}-\hat X_{n, i}}
    {\hat \sigma_{n, i}\sqrt{1 +L_i (M'_{(i)} M_{(i)})^{-1}L_i'}} \, .
$$
It is easy to see that the $\hat e_i(X)$'s are identically distributed (but not independent), with common distribution $St(n-p-1)$. In fact, one can check that the $\hat e_i(X)$'s are the so-called ``Studentized residuals'' of Model \eqref{mod} (see for instance \cite{Cornillon2011book}, Exercise 4.4 p. 87).

We see that the distribution of the random variable $\hat e_i(X)$ does not depend on the unknown parameters $(\theta, \sigma^2)$. In fact, we shall prove in Proposition \ref{main} a much stronger result: the distribution of the random vector $(\hat e_1(X), \ldots, \hat e_n(X))'$ does not dependent on $(\theta, \sigma^2)$ (but it depends on the design $M$). This important fact justifies the procedure below.

To see if there is at least one abnormal observation in the sequence $x_1, \ldots , x_n$, we consider the statistic
$$
    T_n= \max_{i \in \{1, \ldots , n\}} \left | \frac {X_i-\hat X_{n,i}}
    {\hat \sigma_{n, i}\sqrt{1 +L_i (M'_{(i)} M_{(i)})^{-1}L_i'}} \right |= \max_{i \in \{1, \ldots , n\}} \left |\hat e_i(X) \right | \, .
$$
From the discussion of the previous paragraph, we deduce that the distribution of the statistic $T_n$ does not depend on $(\theta, \sigma^2)$. Hence, for each design matrix $M$, it can be tabulated via a basic Monte-Carlo procedure. For a given risk level $\alpha \in (0,1)$, we say that there is at least one abnormal observation in the sequence $x_1, \ldots , x_n$ if the observed value $t_n$ of $T_n$ is such that $t_n>c_{\alpha,n}$, where  $c_{\alpha,n}$ is the quantile of order $1-\alpha$ of the distribution of $T_n$. As usual, this is equivalent to the following procedure: for a given risk level $\alpha \in (0,1)$, we say that there is at least one abnormal observation in the sequence $x_1, \ldots , x_n$ if the $p$-value $p_n={\mathbb P}(T_n>t_n)$ is less than the risk $\alpha$.

It is not possible to provide a table of quantiles of $T_n$ for each possible design matrix $M$. The solution is then, for a given risk level $\alpha$ and a given design matrix $M$, to estimate via Monte-Carlo the quantile $c_{\alpha,n}$ and/or the $p$-value $p_n$. This is exactly what the package \textbf{DEViaN-LM} does.

Let us now give the main result of this section. Consider Model \eqref{mod}, and define the vector $Y$ by
$$
Y=  \frac{X-M\theta}{\sigma} \, ,
$$
in such a way that the coordinates $(Y_i)_{1\leq i \leq n}$ are independent and identically distributed, with ${\mathcal N}(0,1)$ marginal distribution. Let also
$$
 \hat \theta_{n, i}(Y)= \left(M_{(i)}' M_{(i)}\right)^{-1} M'_{(i)} Y_{(i)}\, , \quad
 \hat \sigma_{n, i}^2(Y)= \frac 1 {n-p-1}
 \sum_{k \in \{1, \dots, n\}, k \neq i}
 \left(Y_k-L_k \hat \theta_{n, i}(Y)\right)^2 \, ,
$$
and
$$
 \hat Y_{n, i}= L_i  \hat \theta_{n, i}(Y), \quad
 \hat e_i(Y)=\frac {Y_{i}-\hat Y_{n, i}}
{\hat \sigma_{n, i}(Y)\sqrt{1 +L_i (M'_{(i)} M_{(i)})^{-1}L_i'}} \, .
$$

\begin{proposition}\label{main} In Model \eqref{mod}, the following equality holds: for any $i \in \{1, \ldots, n\}$
$$
  \hat e_i(X)=\hat e_i(Y) \, .
$$
Consequently, the distribution of $(\hat e_1(X), \ldots, \hat e_n(X))'$ does not dependent on $(\theta, \sigma^2)$.
\end{proposition}

\begin{remark}
    If we do not assume that the design matrix $M$ is fixed, then the last assertion of Proposition \ref{main} writes as follows:
    the conditional distribution of $(\hat e_1(X), \ldots, \hat e_n(X))'$ given $M$  does not dependent on $(\theta, \sigma^2)$.
\end{remark}

\begin{remark}
The statistics $T_n$ is the sup-norm of the random vector $(\hat e_1(X), \ldots, \hat e_n(X))'$. From Proposition \ref{main}, we see that the $\ell_p$-norm ($p\geq 1$) of $(\hat e_1(X), \ldots, \hat e_n(X))'$ is also distribution-free, and could be used  to detect if the series $x_1, \ldots, x_n$ is abnormal. The advantage of $T_n$ is the following: once we have computed the quantile  $c_{\alpha,n}$ from the distribution of $T_n$, then one can exhibit all the abnormal observations (if any) at level $\alpha$, that is the observations $x_i$ whose residuals $|\hat e_i(x)|$ are above the threshold $c_{\alpha,n}$.
\end{remark}

{\bf Proof of Proposition \ref{main}.} By definition of $Y_i$ and $\hat Y_{n, i}$, one can write
\begin{align}
Y_i-\hat Y_{n,i}&= \frac{X_i}{\sigma}- \frac{L_i \theta}{\sigma} -L_i \left(M_{(i)}' M_{(i)}\right)^{-1} M'_{(i)} \frac{X_i}{\sigma}
+L_i \left(M_{(i)}' M_{(i)}\right)^{-1} M'_{(i)} M_{(i)}\frac{\theta}{\sigma} \nonumber \\
&= \frac{X_i}{\sigma} -L_i \left(M_{(i)}' M_{(i)}\right)^{-1} M'_{(i)} \frac{X_i}{\sigma} \nonumber \\
&=\frac{X_i-\hat X_{n,i}}{\sigma} \label{one}
\end{align}
In the same way
$$
    Y_k-L_k \hat \theta_{n, i}(Y)=\frac{X_k-L_k \hat \theta_{n, i}}{\sigma}\, ,
$$
and consequently
\begin{equation}\label{two}
    \hat \sigma_{n, i}^2(Y)= \frac{ \hat \sigma_{n, i}^2}{\sigma^2} \, .
\end{equation}
The result follows from \eqref{one}, \eqref{two} and the definition of $\hat e_i(X)$ and $\hat e_i(Y)$.  $\square$

\section{Applications to biological and sociodemographic data}
The package \textbf{DEViaN-LM} provides a convenient way to detect the values poorly explained by a Gaussian linear model or, in other words, to detect abnormal values in the Gaussian framework. As described above, the package performs a series of simple operations in order to estimate the quantiles of $T_n$ via Monte-Carlo simulations for a given model design and level $\alpha$. The abnormal observations at level $\alpha$ can then be found. The package takes 5 input parameters: the variable of interest and the explanatory variables, the quantile of interest, the total number of simulations for estimating the quantiles and the number of CPU to use for the simulations. For example, in order to estimate the grip strength \textit{df\$grip\_strength} of mouse lemurs with regards to their age \textit{df\$age} and body mass \textit{df\$body\_mass} using the following model in \textbf{R}: \textit{df\$grip\_strength \(\sim \) df\$age + df\$body\_mass}, the package \textbf{DEViaN-LM} is called using the following code:
\lstset{style=Rstyle}
\begin{lstlisting}
R> y = df$grip_strength
R> x = cbind( 1, df$age, df$body_mass )
R> model = devianlm_stats(y, x, n_sims = 1e5, quant = 0.95)
\end{lstlisting}
where the default input values are set to $10^5$ simulations and one processor. The user needs to build the model design and send it to the package. For instance, for a 2nd order polynomial regression, one can typically use:
\begin{lstlisting}
R> x <- rnorm(10)
R> e <- rnorm(10, 0, 0.5)
R> y <- 0.5 + 2 * x + x^2 + e
R> model = devianlm_stats(y, cbind( 1, x, x^2 ))
\end{lstlisting}
Note that \textit{devianlm\_stats()} does not add an intercept in the design. If an intercept is needed (as in the preceding examples) the user has to include a column of 1 in the design.

The package returns 4 objects: a (double) vector containing the studentized residuals $\hat e_i(X)$, a (integer) vector with the indices (positions in the original data) of observations identified as outliers based on the threshold, the estimated threshold value $c_{\alpha,n}$ and a (integer) binary vector (0 or 1) indicating whether each observation is considered an outlier (1) or not (0). We illustrate the use of the \textbf{DEViaN-LM} in 4 different datasets: longitudinal sequences of biological markers in Elite soccer players, wage of a sample of the US population, grip-strength in a mouse lemur cohort and wheel activity in a cohort of mice. In each of these datasets, we set the number of simulations to $10^5$ simulations.

\subsection{Elite soccer players}
The description of the original dataset is provided in \cite{sauliere2019z}. New samples were added and the new dataset now consists in a database of 5 typical biological markers from 3936 male soccer players from the French elite leagues 1 \& 2. These biological markers include concentrations of ferritin (mol/L), serum iron ($\mu$mol/L), hemoglobin (g/L), erythrocytes (T/L) and hematocrit levels (\%). They were collected every 6 months in July/August and in January/March from 2005 to 2017. The smallest interval between two measures is 6 months, allowing for independent sampling (\cite{sharpe2006third}). The procedure for selecting and testing the data sample is identical as in \cite{sauliere2019z}. We use \textbf{DEViaN-LM} for assessing the number of abnormal values of serum iron in the studied population.
\subsection{US wage sample}
\label{sec:wages}
The Current Population Survey (CPS) is used to supplement census information between census years. This data consist of a random sample of 599 persons from the 2012 CPS, with information on wages and other characteristics of the workers. We use \textbf{DEViaN-LM} for assessing abnormal values of the logarithm of the salary / wage, adjusted to the age, the educational level and the number of children.
The linear model is: $\log(\text{salary}) = \beta_0 + \beta_1 \text{age} + \beta_2 \text{educational level} + \beta_3 \text{nb of children}$
\subsection{Grip strength of mouse lemurs}
The maximal grip strength of mouse lemurs (\textit{Microcebus murinus}) was measured using a small iron bar mounted on a piezo-electric force platform (Kistler squirrel force plate, $\pm0.1$N; Winterthur, Switzerland), connected to a charge amplifier (Kistler charge amplifier type 9865). The pull strength was recorded during 60s at 1kHz. Animals performed the pull task several times during this interval. The task consisted of  letting the animal grab the iron bar and pull them horizontally until they let it go. The maximum force was then extracted using the Bioware software (Kistler). The repeatability of this task was previously verified (\cite{thomas2016determinants}). A sample of 359 individuals was measured: 170 females and 189 males, with 39 individuals tested twice. A total of 397 grip-strengths (N) were measured. We use the \textbf{DEViaN-LM} for assessing the number of abnormal (log) values in the grip-strengths, in regards to age and body mass. The dataset consists in 340 grip-strengths, with corresponding age and body mass (57 body mass values are missing). See above for the equation used.
\subsection{Wheel activity in mice}
The maximal distance per day in the wheel activity is recorded for mice (\textit{Musdomesticus}) based on their voluntary behavior to practice wheel activity (\cite{morgan2003ontogenies}). A total of 159 mice were genetically selected for high locomotor activity and 14241 daily wheel running performances (7078 for males and 7163 for females) were gathered (\cite{bronikowski2006evolution}). The distance run was converted to km/week and we use the \textbf{DEViaN-LM} for assessing abnormal distances with regards to the age and body mass. We select the first lineage, as the wheel running distances of this lineage appear more normally distributed compared to the distances of other lineages. The dataset consists in 1673 wheel running distances from 19 mice (9 females and 10 males) with corresponding age and body mass.

\begin{figure}
\centering
\includegraphics[width=1\textwidth]{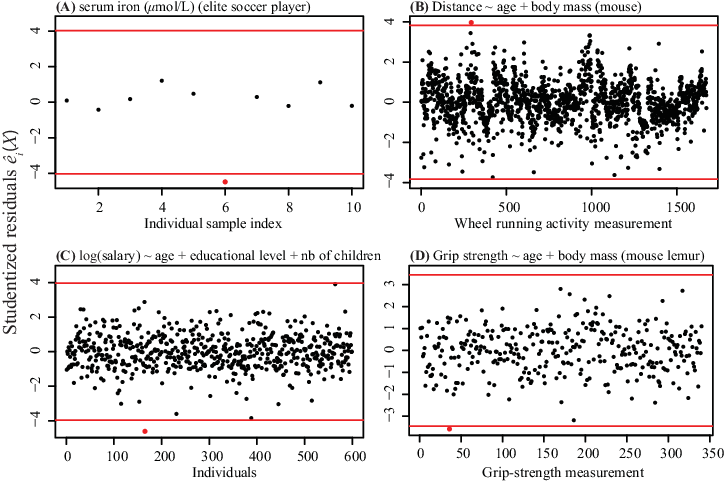}
\caption{Application of \textbf{DEViaN-LM} to biological data in 4 examples. Black dots represent the data points, red dots show abnormal values, and red lines represent the quantiles for the requested $\alpha$ level (for all pictures we took $\alpha= 5 \%$). The upper left panel illustrates the detection of abnormal values in the serum iron samples for one elite soccer player (\textbf{A}). In the upper right panel (\textbf{B}); the abnormal values of distance per week, adjusted by age and bodymass (weight) are given for mice. The lower left panel (\textbf{C}) reveals abnormal salaries adjusted by age, educational level and the number of children. The lower right panel (\textbf{D}) shows the mouse lemurs which present abnormal grip strength values, when adjusting for age and weight.}
\label{fig:Fig1}
\end{figure}

\section{Software and benchmark}
The package performs a series of simple operations in order to estimate the quantiles of $T_n$ via Monte-Carlo simulations for a given model design and level $\alpha$. The precision of these quantiles estimates increases with the number of simulations performed in the Monte-Carlo process. However, increasing the number of simulations can be both time- and resource-intensive. In this section we detail the typical methods for computing the quantiles of $T_n$ using Monte-Carlo simulations in \textbf{R} and \textbf{MATLAB}. We compare their running times with a custom \textbf{C++} routine, developed for the proposed package.
\subsection{R and MATLAB implementations}
\label{sec:RMatlabcode}
The quantiles of the statistic $T_n$ can be computed using the following straightforward \textbf{R} code example:
\begin{lstlisting}
R> for(i in 1:nsimul)
R> {
R>  x <- rnorm(n)
R>  reg2 <- lm(x ~ X)
R>  Max[i] <- max(abs(rstudent(reg2)))
R> }
R> Qu <- quantile(Max, 0.95)
\end{lstlisting}
where \textit{X} is the variable in the model design (such as \textit{X = AGE} in \textit{WAGE ~ AGE}, see section \ref{sec:wages}) and \textit{nsimul} is the number of simulations to perform for estimating the quantiles. Similarly, the following \textbf{MATLAB} code can be used to compute the quantiles of $T_n$:
\begin{lstlisting}[style=Matlab-editor]
MATLAB> for i=1:nsimul
MATLAB>  x = randn(s, 1);
MATLAB>  mdl = fitlm(X, x);
MATLAB>  Max(i) = max(abs(mdl.Residuals.Studentized));
MATLAB> end
MATLAB> Qu = quantile(Max, 0.95);
\end{lstlisting}
However, in both cases, the usage of \textit{lm()} or \textit{fitlm()} in the \textit{for()} loop is time consuming when the number of simulations \textit{nsimul} is large. The \textit{lm()} and \textit{fitlm()} functions compute many additional statistics that are not required for computing $T_n$. In contrast, \textbf{DEViaN-LM} only computes the (externaly) studentized residuals, resulting in lesser operations overall.

\subsection{Benchmark}
\label{sec:Benchmark}
The efficiency or performance of the \textbf{DEViaN-LM} package in terms of runtime was assessed using a simple benchmark procedure. It was carried out on a desktop computer equipped with 6 INTEL Xeon E-2186G CPU at 3.80GHz, 16GB of RAM (2666MHz) and running \textbf{R} 4.3.2 and \textbf{MATLAB} 9.3.0.713579 (R2017b). For the \textbf{R} environment, runtime measurements were obtained using the \textit{microbenchmark} package \cite{mersmann2019microbenchmark}, while the \textit{tic-toc} functionality was employed for \textbf{MATLAB}. The performance was compared to a typical \textbf{R} and \textbf{MATLAB} code implementation (see previous section \ref{sec:RMatlabcode}). Two effects were measured: the sample-size effect and the number of Monte-Carlo simulations. In the first case, a dataset with a large number of observations was simulated. In the second case, the number of simulations was increased. In order to have a consistent measure of the time taken to run \textit{nsimul} simulations , we repeated each measure and took the median value of the runtime (in seconds). For the sampling-size effect, we repeated each measure 200 times and we used a simulated linear model $y_i = 25.0 + 3.4x_i + \epsilon_i$ with $x \sim \mathcal{N}(0,\,1)$, $\epsilon \sim \mathcal{N}(0,\,2)$ and $i = \left\{ 100,\, 500,\, 10^3,\, 5\times10^3,\, 10^4,\, 10^5,\, 10^6 \right\}$. For the Monte-Carlo simulations we repeated each measure 100 times and used the WAGE dataset (section \ref{sec:wages}), with \textit{nsimul} taking the following values: $\text{nsimul} = \left\{ 100,\, 500,\, 10^3,\, 5\times10^3,\, 10^4,\, 1.5\times10^4,\, 2.0\times10^4,\, 2.5\times10^4 \right\}$.

\begin{figure}
\centering
\includegraphics[width=1\textwidth]{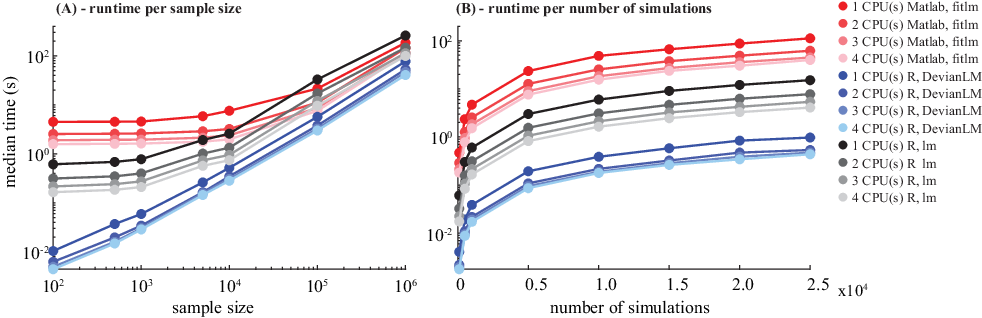}
\caption{Benchmark results. In the left panel (\textbf{A}), the median runtime is given for different sample-sizes and different number of CPU cores. Each measure (node) is repeated 200 times. Similarly, in the right panel (\textbf{B}), the median runtime is given for an increasing number of simulations \textit{nsimul} using $100\%$ of the WAGE dataset. Each measure is repeated 100 times.}
\label{fig:Fig2_benchmarck}
\end{figure}

Results are provided in Figure (Fig. \ref{fig:Fig2_benchmarck}) and show that the median runtime increases with the sampling-size and the number of simulations \textit{nsimul} (Fig. \ref{fig:Fig2_benchmarck}). The increase in the runtime can be estimated using a linear regression of the form $y = rx +b$ where $r$ denotes the rate of increase (Table \ref{tab:linReg}). The package \textbf{DEViaN-LM} has the smallest $r$ in both cases (Table \ref{tab:linReg}).

\begin{table}
\centering
\begin{tabular}{lllccccc}
\hline
type & language & method & \textit{r} & t-stat & \textit{p}-value & RMSE & R2a\\ \hline
sample-size & \textbf{MATLAB} & \textit{fitlm()} &  $1.80{\scriptstyle\times10^{-4}}$ & 212.91   & $0.77$ & 0.18  & 0.99\\
sample-size & \textbf{R} & \textit{lm()} &          $2.58{\scriptstyle\times10^{-4}}$ & 86.84   & $2.71$ & 0.048 & 0.99\\
sample-size & \textbf{R} & \textbf{DEViaN-LM} &   $7.72{\scriptstyle\times10^{-5}}$ & 86.45  & $0.82$ & 0.00087& 0.99\\ \hline
sim. number & \textbf{MATLAB} & \textit{fitlm()} &  $4.43{\scriptstyle\times10^{-3}}$ & 64.11   & $9.68{\scriptstyle\times10^{-10}}$ & 1.75  & 0.99\\
sim. number & \textbf{R} & \textit{lm()} &          $6.06{\scriptstyle\times10^{-4}}$ & 493.16  & $4.69{\scriptstyle\times10^{-15}}$ & 0.031 & 0.99\\
sim. number & \textbf{R} & \textbf{DEViaN-LM} &   $3.99{\scriptstyle\times10^{-5}}$ & 47.06   & $6.17{\scriptstyle\times10^{-9}}$  & 0.022 & 0.99\\ \hline
\end{tabular}
\caption{\label{tab:linReg} Estimates of growth rate $r$ using linear regression for different samples-sizes and number of simulations for 1 CPU. The value of $r$ is given in seconds per sample-size (for samples-sizes, Fig. \ref{fig:Fig2_benchmarck}, left panel) and in seconds per number of simulations (for different number of simulations, Fig. \ref{fig:Fig2_benchmarck}, right panel). The t-stat is the t-statistic for $r$ which is the slope of the regression line and \textit{p}-value is the associated $p$ value. The RMSE and adjusted R\textsuperscript{2} are given to assess the quality of the linear regression.}
\end{table}

\section{Discussion and concluding remarks}
One statistical approach when searching for abnormal values or outliers in a linear model relies on the analyse of the studentized residuals. Studentized residuals quantify how large the residuals are in standard deviation units, and therefore can be used to identify outliers: An observation with an externally studentized residual that is larger than 2 (in absolute value) is generally deemed an outlier (see for instance Section 4.1 and paragraph 4.1.2 in \cite{Cornillon2011book}). They are typically used to assess the potential outliers that influence a regression model to such an extent that the estimated regression function is attracted towards the potential outlier. The usual way to proceed is to estimate the studentized residuals then look at which ones seem "off the chart" so to speak.

Note that looking at whether each studentized residual is greater than the threshold (calculated from Student's distribution) gives rise to a multiple testing problem. To be convinced of this, suppose that we observe $n$ values $x_1, \ldots , x_n$ obtained from a sample $X_1, \ldots , X_n$ distributed according to the ${\mathcal N}(0,1)$ law. If we say that a value is abnormally large if its absolute value exceeds the threshold of 1.96, then we know that, if $n=100$, we should observe between 4 and 6 "abnormal" values. These "abnormal" values are in fact perfectly normal, and simply come from the repetition of draws according to the ${\mathcal N}(0,1)$ distribution. We are here exactly in the issue of multiple tests. One way to correct this issue is to identify the observations whose absolute values are above the threshold calculated from the law of $\max_{1\leq i \leq n}|X_i|$. These values can really be considered abnormally large.

The method we propose follows exactly this approach, in the context of the Gaussian linear model. It computes the exact distribution of the maximum values of the studentized residuals, based on the design of the model. The estimate of the quantiles can only be computed using numerical simulations as the quantiles depends on each design. The \textbf{DEViaN-LM} package automates this process and provides users with a unique and practical tool for measuring abnormal values in the Gaussian Framework. Its implementation is efficient and compares very well with different build-in functions (Fig. \ref{fig:Fig2_benchmarck}). We focused on the free \textbf{R} software environment, but the proposed \textbf{C++} implementation can be easily adapted to other languages.

\section{Acknowledgments}
We thank Brigitte Gelein, Thibaud Lef\`{e}vre, Kenza Charifi and Aliya Mohamed Soultane for providing additional help in coding the package.

\bibliographystyle{unsrt}
\bibliography{bibfile_Zscore}

\end{document}